\documentclass[useAMS,usenatbib]{mn2e}

\usepackage{graphicx}
\newcommand{\lcdm}{$\Lambda$CDM }
\newcommand{\de}{{\mathrm d}}
\newcommand{\mhi}{M_{H\! I} }
\newcommand{\mstar}{M_{\ast} }
\newcommand{\msun}{M_{\odot} }
\newcommand{\hi}{$HI$ }
\newcommand{\hii}{$H_{2}$ }
\newcommand{\lsun}{$L_{\odot}$ }
\newcommand{\msolar}{$M_{\odot}$ }
\newcommand{\lstar}{$L_{\ast}$ }

\title[Two phase galaxy formation: The Gas Content of Normal Galaxies]{Two phase galaxy formation: The Gas Content of Normal Galaxies}

\author[M. Cook et al.]{M. Cook$^{1,2}$\thanks{E-mail:cook@sissa.it (MC)}, C. Evoli$^{1}$, E. Barausse$^{5,1}$, G.L. Granato$^{4,2}$, A. Lapi$^{3,1,4}$\\
$^{1}$Astrophysics Sector, SISSA/ISAS, Via Beirut 2-4, I-34014 Trieste, Italy\\
$^{2}$INAF, Osservatorio Astronomico di Padova, Vicolo dell' Osservatorio 5, I-35122 Padova, Italy\\
$^{3}$Dept. of Physics, Univ. di Roma `Tor Vergata', Via della Ricerca Scientifica 1, I-00133 Rome, Italy\\
$^{4}$INAF, Osservatorio Astronomico di Trieste, Via G.B. Tiepolo 11, I-34131 Trieste, Italy\\
$^{5}$Centre for Fundamental Physics, University of Maryland, College Park, MD 20742-4111, USA}

\begin{document}

\maketitle

\label{firstpage}

\begin{abstract}

We investigate the atomic ($HI$) and molecular ($H_{2}$) Hydrogen content of normal galaxies by combining observational studies linking galaxy stellar and gas budgets to their host dark matter (DM) properties, with a physically grounded galaxy formation model. This enables us to analyse empirical relationships between the virial, stellar, and gaseous masses of galaxies and explore their physical origins. Utilising a semi-analytic model (SAM) to study the evolution of baryonic material within evolving DM halos, we study the effects of baryonic infall and various star formation and feedback mechanisms on the properties of formed galaxies using the most up-to-date physical recipes. We find that in order to significantly improve agreement with observations of low-mass galaxies we must suppress the infall of baryonic material and exploit a two-phase interstellar medium (ISM), where the ratio of \hi to \hii is determined by the galactic disk structure. Modifying the standard Schmidt-Kennicutt star formation law, which acts upon the total cold gas in galaxy discs and includes a critical density threshold, and employing a star formation law which correlates with the \hii gas mass results in a lower overall star formation rate. This in turn, allows us to simultaneously reproduce stellar, \hi and \hii mass functions of normal galaxies.

\end{abstract}


\section[]{Introduction}

Neutral atomic hydrogen is the most abundant element in the Universe and plays a fundamental role in galaxy formation, principally as the
raw material from which stars form. Within galaxies, the Interstellar Medium (ISM) acts as  a temporally evolving baryonic component; competing processes cause the accumulation (through external infall from the intergalactic medium and stellar evolution) and depletion (through star formation and various feedback mechanisms) of hydrogen. Thus, observational determinations and theoretical predictions of the hydrogen budget within galaxies of
various masses and morphologies is of central importance to constraining the physics of galaxy formation (see Kauffmann, White \& Guideroni, 1993, Benson et al. 2003, Yang, Mo \& van den Bosch, 2003, Mo et al. 2005, Kaufmann et al. 2009)

Moreover, within the ISM Hydrogen comprises the majority of the cold gas mass, and when non-ionized exists within two-phases, atomic \hi and molecular \hii. A large body of observational analysis has shown that within galaxies, \hi generally follows a smooth, diffuse distribution whereas \hii regions are typically dense, optically thick clouds which act as the birthplaces for newly formed stars (Drapatz \& Zinnecker, 1984, Wong \& Blitz, 2002, Krumholz \& McKee, 2005, Blitz \& Rosolowski, 2004, 2006, Wu et al., 2005). Due to the distinct differences in these phases, and the central importance of ISM physics to the evolution of galaxies, cosmological simulations have begun to include both phases (see Gnedin et al. 2009 \& references therein), and observations have begun focusing on simultaneous measurements of both \hi and \hii (see Obreschkow \& Rawlins, 2009).

The distinction between these two phases has recently been shown to be of crucial importance to constrain the physics of galaxy formation.  in particular resolved spectroscopy using  GALEX showing obscured star forming regions in nearby galaxies (Kennicutt et al. 2003, 2007, Calzetti et al. 2007, Gil de Paz et al. 2007), and various observational surveys providing maps of gas in galaxies at high-resolutions (Walter et al. 2008, Helfer et al. 2003, Leroy et al. 2009), have revealed a deeper level of complexity on sub-galactic scales. These studies allowed theoretical models for the ISM and star formation to be constrained and further developed.

\begin{figure}
\includegraphics[width=84mm]{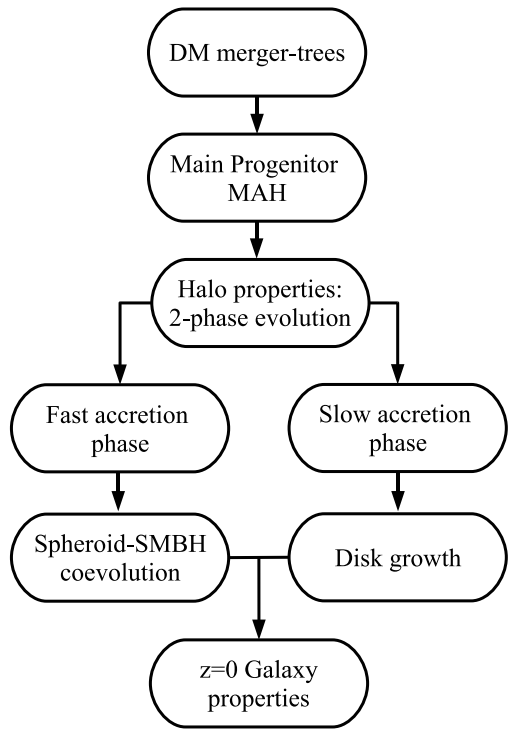}
\caption{Schematic of the model framework, showing how we partition the evolutionary history into spheroid and disc growth epochs, which follow significantly different evolutionary paths for baryonic structure growth and result in the bulge-disk dichotomy observed at $z=0$.}
\label{schematic}
\end{figure}

Furthermore, due to the constant replenishment and depletion of Hydrogen in either \hi or \hii phases, and to their separate yet interlinked properties, at any epoch, measurements of the fraction of \hi and \hii are highly constraining for the processes of molecular cloud formation, star formation, baryonic infall and various feedbacks. Therefore, simultaneously predicting the stellar and gas mass functions of normal galaxies is a major challenge for any physically motivated galaxy formation model, requiring an accurate depiction of all of the aforementioned processes (see Mo et al. 2005 for a detailed discussion). These issues manifest most clearly within the largely successful \lcdm paradigm within the lowest mass systems, where it still remains unclear whether strongly non-linear feedback mechanisms, lower star formation efficiencies, or suppression of initial infall onto DM halos is the dominant driver for the suppression of luminous structure formation (Mo et al. 2005). It is more than likely that a combination of the above-mentioned effects will go a long way to alleviating current tensions between models and observations, since current semi analytical models (SAMs) incorporate several processes in order to generate a deficiency of stellar mass in DM halos; many of which operate most effectively at low masses (Benson et al. 2003, De Lucia et al., 2004).

Observationally, Zwaan et al. (2005) used the catalogue of $4315$ extragalactic \hi 21-cm emission line detections from the \hi Parkes All Sky Survey (HIPASS, Barnes et al. 2001) and obtained the most accurate measurement of the \hi mass function of galaxies to date. The HIMF is fitted with a Schechter function with a faint-end slope of $-1.37 \pm 0.03$. The sensitivity of this survey was so high that they were able to extend their analysis well down to \hi masses of $10^{7.2} M_\odot$, hence this is most complete analysis so far. Using these statistical constraints, it has now become possible to make stringent comparisons between theoretical models and observations even in low mass galaxies.

The physics of cold gas becomes increasingly relevant for constraining galaxy formation models at relatively low masses (dominated by late-type galaxies), where the presence of gas becomes substantial and therefore may break the degeneracies between feedback, star formation, and infall processes. Moreover, within the \lcdm scenario the \hi and \hii mass budgets in galaxies are determined by an intricate offset between several competing processes, all of which have strong mass dependencies. Thus the present \hi and \hii fractions are strong functions of host DM halo mass and the evolutionary history of each individual galaxy. More specifically, the fraction of gas which may be captured by the host DM halo, and in turn removed by feedback, is expected to depend strongly on the binding energy of the gas itself, which is principally determined by the DM halo virial mass and density distribution. Thus, under this framework the observational properties of galaxy populations are strongly influenced by their collective host DM halos (White \& Rees, 1978, see Somerville et al. 2008 for a review).

Motivated by the above-mentioned observational advances and theoretical challenges, the primary aim of this work is to investigate the physical origins of the relationships between \hi, stellar, and virial masses of galaxies. In order to do this we compare empirical galaxy relations derived from observational studies, with a physically motivated galaxy formation model. Observationally, we use a numerical approach (described in Shankar et al. 2006) that relies on the assumption of the existence of a one-to-one mapping between galaxy properties and host DM halo mass\footnote{This approach is based also on the assumption on the completeness of the sample over which the mass functions has been obtained.}. We interpret these results using a physically motivated SAM, (see Cook et al. 2009, hereafter C09, \& references therein), which has been shown to reproduce many features of the local galaxy population.

SAMs provide a powerful theoretical framework within which we can explore the range of physical processes (e.g. accretion mechanisms, star formation, SN feedback, black hole growth and feedback etc.) that drive the formation and evolution of galaxies and determine their observable properties (see Somerville et al. 2008 and Baugh 2006 for extensive reviews) whilst remaining computationally feasible, allowing for statistical samples of galaxies to be generated.

As is shown in Fig.1, the backbone to our galaxy formation model is a cosmologically consistent DM halo merger-tree, outlining the merging and accretion histories of DM halos. For the evolution of baryonic structures within these merging trees, we utilize the model of C09, with several modifications. This requires us to partition each mass accretion history (MAH) into two phases: A \emph{'fast accretion'}, merger dominated phase corresponding to spheroid-SMBH co-evolution, followed by a \emph{'slow accretion'}, quiescent phase allowing for disk structure to form around the pre-processed halos (see also Zhao et al. 2003, Mo \& Mao, 2004), the resulting galaxy properties at $z=0$ are thus a result of the sub-grid baryonic recipes, and the mass accretion histories of their host DM halos. Within this work we also expand the previous model by including several additional effects which are thought to be of crucial importance the low-mass gas-rich systems, namely, we model the effects of an ionizing ultraviolet (UV) background, cold accretion flows and exploit a two-phase ISM, where star formation is governed by the surface density of \hii.

In summary, we utilise the most up-to-date observations of the stellar and \hi mass functions in order to derive relationships between the host DM halo and the galaxy stellar and gas properties, we interpret these results using a physically grounded SAM, and analyse the nature of the results by generating three model realisations, incorporating several contemporary recipes and highlighting how each helps alleviate previous model tensions.  The outline of the paper is as follows: In \S2 we present our SAM, outlining the basic framework from C09, and including the modifications and improvements, in \S3 we use current observational results of the stellar and \hi mass functions and use a numerical procedure in order to derive relationships between host DM halo mass and the galaxy stellar and \hi components. In \S4 We present the results of both the observational determinations and the theoretical model comparisons. We conclude in \S5 by summarising the main outcomes of this work, discussing the implications, and the limitations.

Throughout the paper we adopt the standard $\Lambda$CDM
concordance cosmology, as constrained by \textsl{WMAP} 5-year
data (Spergel et al. 2007). Specifically, we adopt a flat
cosmology with density parameters $\Omega_M=0.27$ and
$\Omega_{\Lambda}=0.73$, and a Hubble constant $H_0=70$ km
s$^{-1}$ Mpc$^{-1}$.

\begin{figure}
\includegraphics[height=85mm, angle=90]{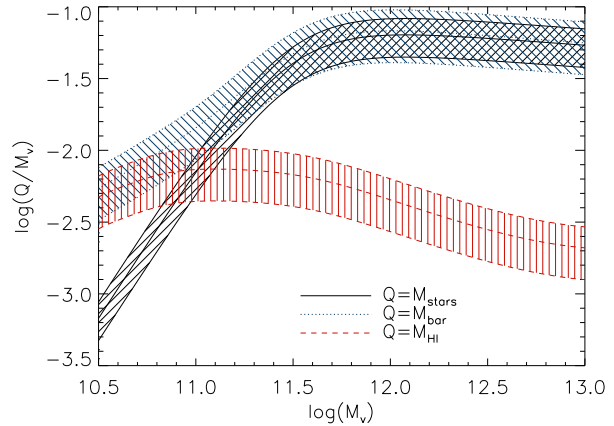}
\caption{The derived relationships between virial mass and; stellar (solid/black), \hi (dashed/red) and baryonic (dotted/blue) components. Shaded regions corresponding to the $1-\sigma$ gaussian errors associated with the 
observational determinations.}
\label{}
\end{figure}

\begin{figure*}
\begin{center}
\includegraphics[height=185mm, angle=90]{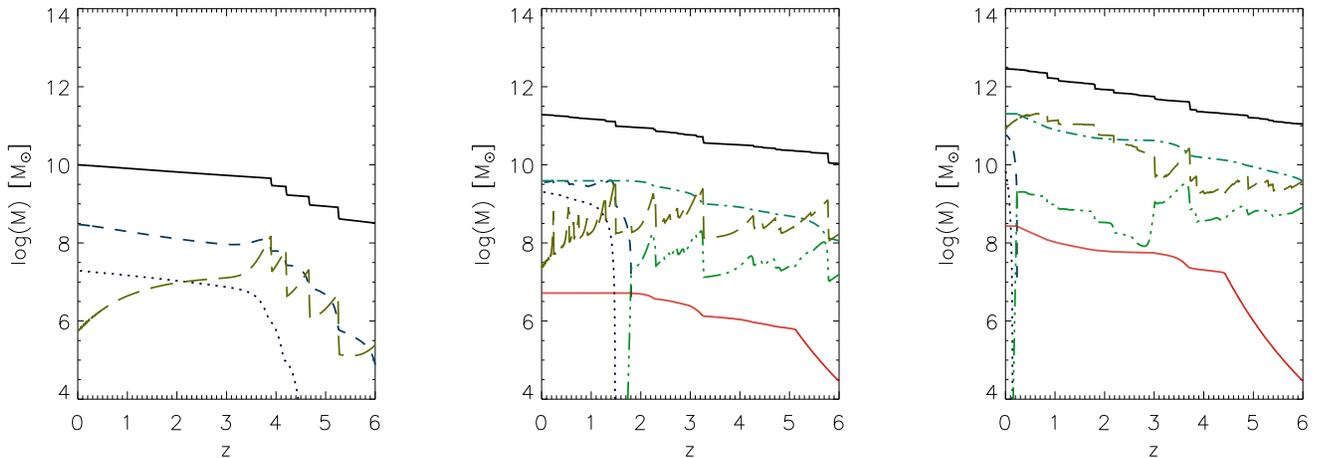}
\caption{The evolution of typical model galaxies; a low-mass disk-dominated galaxy (left-hand panel), an $L_{\odot}$ galaxy with both bulge
and disks present (middle panel), and a high mass, spheroid dominated galaxy (right panel): DM mass (solid/black), hot gas (long-dash/brown), disc gas (short-dash/blue), disc stars (dotted/purple), bulge gas (triple-dot-dashed/green), bulge stars (dot-dashed/aquamarine), BH mass (solid/red)}
\label{MW}
\end{center}
\end{figure*}

\section[]{Physical model}


We follow the galaxy evolution using a semi-analytical approach, deriving analytical
estimates wherever possible to aid simplicity. We track the evolution of dark matter,
hot gas, cold gas and stellar mass using several parameterizations, based on the model described in
C09 but with several important modifications. 

Namely, we now  trace the sequential infall of material onto both spheroid and disc structures and we include the effects of an ionizing UV background, adiabatic contraction of the DM halo and a two-phase ISM. Here we outline the basic framework for this model.

We initially describe the dark matter halo evolution using an extended Press-Schechter (EPS) algorithm based
on that developed by Parkinson et al. (2008). These mergertrees have been tuned in order to reproduce the
statistics of halo merger and accretion activity obtained from N-body simulations of structure formation (Springel et al.
2005). Within our model, we utilise these mergertrees by extracting the mass accretion history (MAH), \emph{i.e.} the
evolutionary path of a typical dark matter halo, obtained by tracking the most massive progenitor at
each fragmentation event whilst moving from $z=0$ to progressively higher redshifts (see van den Bosch,
2002, C09).

Specifying the DM halo properties, we define the virial radius $r_{\rm vir}$ as that of a spherical volume enclosing an over-density in a standard way (Bryan \& Norman, 1998). We compute the density profile for the dark matter halo
using the fitting function of Navarro, Frenk \& White (1997, hereafter NFW):
\begin{equation}\label{profile}
    \rho_{_{\rm NFW}}(r) = \rho_{s} \left(\frac{r}{r_{s}}\right)^{-1} \left(1+\frac{r}{r_{s}}\right)^{-2}\,.
\end{equation}
In order to define the scale radius $r_s$ for our NFW profile  we introduce the \emph{concentration parameter}, which
is defined to be $c(z) \equiv r_{\rm vir}/r_{s}$. This quantity has been studied by several authors [see Bullock et al. 2001,
Zhao et al. 2003 (hereafter Z03), Maccio et al. 2007] who found large scatter for a fixed halo mass, but to scale
generally with the MAH of the halo. We adopt here the $z$-evolution of Z03,
\begin{equation}\label{11}
  { [\ln(1+c)-c/(1+c)]c^{-3\alpha}  }   \propto {H(z)}^{2\alpha} {M_{\rm vir}(z)}^{1-\alpha}\,,
\end{equation}
where $\alpha$ is a piecewise function which can be found in Z03 and where the normalization can be fixed using the expression given by Maccio et al. 2007 at $z=0$:
\begin{equation}
\log c_0 = 1.071 - 0.098\, \left[\log \left({M_{\rm vir,0}\over
M_{\odot}}\right)-12\right]\,.
\end{equation}
From this, the scale density may be computed for a general profile to be $\rho_s=M_{\rm vir}(z) /[4\pi r_{s}^{3} f(c)]$,
with
\begin{equation}\label{fc}
    f(c) = \ln(1+c)-\frac{c}{1+c}\,.
\end{equation}

Finally, we specify the angular momentum properties of each halo through the \emph{spin parameter}, defined to be
$\lambda = J_{\rm vir}E_{\rm vir}^{1/2}M_{\rm vir}^{5/2}G^{-1}$, where $E_{\rm vir}$ and $J_{\rm vir}$ are the total energy and angular momentum of the halo. Assuming that a DM halo acquires its angular momentum through
tidal torques with the surrounding medium, $\lambda$ remains constant. It has been shown (Cole \& Lacey, 1996) that the spin parameter varies little with cosmic epoch, halo mass, or environment and for a sample of haloes is well fitted by a log-normal distribution (where $\bar{\lambda} = 0.04$ and $\sigma =0.5$). However, for simplicity, here we assume that each halo has a parameter value of $\lambda=0.04$ which remains constant throughout the DM halo evolution.

Several studies have shown the concentration evolution to be strongly correlated to the MAH of the halo (Z03,
Li et al., 2006, Lu et al. 2006), finding that DM halos generally acquire their mass in two distinct phases; an initial phase
characterised by rapid halo growth through major merger events, where the halo core structure forms, causing the gravitational
potential to fluctuate rapidly, followed by a slower, more quiescent growth predominantly through accretion of material onto
the outer regions of the halo. These two different modes are reflected in the evolution of the concentration parameter, which remains roughly constant during the 'fast accretion' phase and steadily increases during the 'slow accretion' phase. The transition redshift $z_t$ between these two phases can therefore be calculated using the expression for the concentration parameter evolution given in Z03:
\begin{equation}\label{11}
   \frac{ [\ln(1+c_t)-c_t/(1+c_t)]c^{-3\alpha}  }{ [\ln(1+c_{0})-c_{0}/(1+c_{0})]c_{0}^{-3\alpha}}   = \left[\frac{H(z_t)}{H_{\rm 0}}\right]^{2\alpha} \left[\frac{M_{\rm vir}(z_t)}{M_{\rm vir,0}}\right]^{1-\alpha}\,,
\end{equation}
where $H(z)$ is the Hubble radius, '$0$' denotes quantities evaluated at the present cosmic time, and $c_t=4$ and $\alpha=0.48$ are scaled to match N-body simulations as in Z03.


By associating the two phases of DM evolution to two growth mechanisms for the baryonic sector, the fast and slow phases give rise to the formation of bulges and discs respectively (Mo \& Mao, 2004, C09). Since within the 'fast accretion' phase, angular momentum may be readily lost by rapidly merging clumps of material, in an implicit merger scenario, resulting in the formation of a spheroidal structure, followed by the quiescent dissipationless infall of material in order to form discs (see C09 for a more detailed analysis and justifications of this). 
At difference with the prescriptions outlined in C09, here we expand our model to include the symbiotic infall of baryonic material
as the DM halo evolves, thus
\begin{equation}\label{11}
\dot{M}_{\rm inf}= f_{\rm coll} \dot{M}_{\rm vir}\,.
\end{equation}
We include the effects of an ionizing radiation background taking the prescriptions outlined in (Gnedin et al., 2004, Somerville et al., 2008), which is able to partially reduce the baryonic content in low-mass systems, thus
\begin{equation}\label{1}
   f_{\rm coll}(M_{\rm vir},z) = \frac{\Omega_{b}/\Omega_{m}}{(1+0.26M_{f}(z)/M_{\rm vir})^{3}}\,,
\end{equation}
where $M_{f}(z)$ is the filtering mass at a given redshift, computed using the equations in (Kravtsov, Gnedin \& Kyplin, 2004, Appendix B). A second improvement over our previous model (C09) is to include the effects of cold accretion flows, shown to be the predominant mechanism leading to the formation of low-mass systems. Below a critical mass
\begin{equation}\label{1}
   M_{c} = M_{s} \max[1, 10^{1.3(z-z_{c})}]\,,
\end{equation}
where $M_{s}=2\times10^{12} M_{\odot}$ and $z_{c}=3.2$, we assume that all gas accreted onto DM halos is not shock heated to the virial temperature of the DM halo, but streams in on a dynamical time (see Dekel et al., 2008, 2009, Cattaneo et al, 2006). We note that below the shock heating mass scale it has been shown that rapid cooling does not allow for the formation of a stable virial shock (Keres et al., 2005) resulting in gas flowing unperturbed into the central regions of the DM halo.
Thus, in halos below this mass the collapse happens on the dynamical timescale of the system ($t_{\rm coll} = t_{\rm dyn}$), whereas in halos above this mass $t_{\rm coll} = \max[t_{\rm dyn}, t_{\rm cool}]$, where the cooling timescale $t_{\rm cool}$ is computed in a standard way, assuming material is shock heated to the virial temperature. The effects of this cold accretion is to moderately enhance star formation at high redshifts relative to the scenario where all material is shock heated, in this way, we model the infall and cooling dynamics using the most current recipes (see Somerville et al., 2008).

In order to calculate the cooling time of the hot gas phase, we assume an isothermal gas in hydrostatic equilibrium within the NFW profile, such that
\begin{equation}
    \rho_{\rm hot}(r) = \rho_{0} \exp \left[ \frac{-27}{2} \beta \left\{ 1-\frac{\ln(1+r/r_{s})}{r/r_{s}}   \right\}   \right],
\end{equation}
with
\begin{equation}
   \beta = \frac{ 8 \pi \mu m_{p} G \rho_{s} r_{s}^{2}}{27 k_{B} T_{\rm vir}}\,,
\end{equation}
where $T_{\rm vir}$ is the virial temperature, $\mu$ is the mean molecular mass and $\rho_{0}$ is calculated by normalizing to the total hot gas mass at any given redshift. Also, we assume that the cooling function $\Lambda(T,Z)$ is given by the tabulated function of Sutherland \& Dopita, 1993, and assume that the infalling baryonic material is unprocessed and therefore has primordial metallicity, $Z=10^{-3}Z_{\odot}$.

\begin{table*}
\centering
\caption{Values of the free parameters of our model. We note that the value of $k_{acc}$, despite being a free parameter, has a minor impact on our results}
\begin{tabular}{lcccc}
\hline
\hline
Description &  Symbol & Fiducial value & Reference in the text & Impact on this work\\
\hline
SN feedback efficiency (bulge)         &   $\epsilon_{\rm SN,b}$   &  $0.5$ & Eq.~15 & Strong\\
SN feedback efficiency (disc)        &   $\epsilon_{\rm SN,d}$   &  $0.7$ & Eq.~35 & Strong\\
Reservoir growth rate           &   $A_{res}$               & $10^{-3} M_{\odot}yr^{-1}$   & Eq.~18 & Strong\\
QSO feedback efficiency         &    $f_{h}$                &  $10^{-4}$            & Eq.~22, 23 & Strong\\
Viscous accretion rate & $k_{acc}$ & $10^{-2}$ & Eq.~20 & Weak
\\
\hline
\end{tabular}
\hbox{Note. - A Romano IMF $\phi(m_{\star})$ is adopted:
$\phi(m_{\star})\propto m_{\star}^{-1.25}$ for $m_{\star}\geq
M_{\odot}$ and $\phi(m_{\star})\propto m_{\star}^{-0.4}$ for
$m_{\star}\leq M_{\odot}$.}
\end{table*}

\subsection{ Dissipative gas collapse }

\begin{figure}
\includegraphics[height=85mm, angle=90]{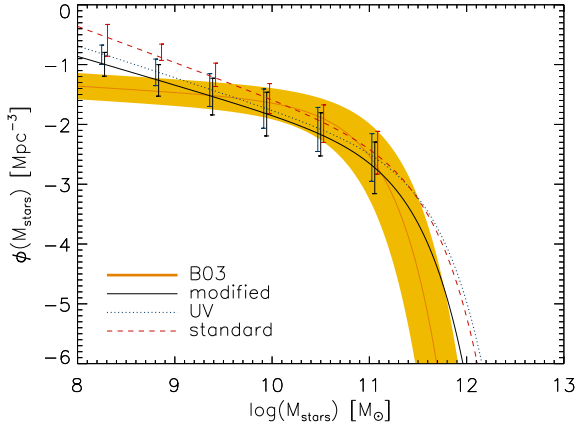}
\caption{Stellar mass function output by each of our three model realisations compared with the Bell et al. 2003a estimate. We find that the most sophisticated treatment of the baryonic physics results in agreement to the observations in all but the lowest mass halos ($M_{s}<10^{9}M_{\odot}$), where we find that we overproduce the stellar matter by approximately a factor two. Note that, due to large uncertanties in observational mass-to-light ratios, we associate 0.3dex errors in the stellar mass determinations, and 0.4dex in the value of $\phi(M_{stars})$, shown by the shaded area. Model error-bars represent poissonian uncertanties due to the finite sample size of synthetic galaxies.}
\label{GSMF}
\end{figure}

During the high redshift domain, it has been shown that the majority of the angular momentum of the collapsing proto-galactic gas dissipates as it collapses and condenses to the centres of DM haloes (see Navarro \& Steinmetz, 2000). Also, within C09 we show that during the early collapse phase, DM halo subunits merge on timescales shorter than the overall dynamical time of the forming halo. Therefore, within the 'fast accretion' phase we neglect the effects of angular momentum of cooling proto-galactic gas, which will result in the formation of a spheroidal gaseous system at a rate
\begin{equation}
    \dot{M}_{\rm coll}(z) = 4 \pi \int_{0}^{r_{\rm vir}(z)} \frac{r^{2} \rho_{\rm hot}(r,z)}{t_{\rm coll}(r,z)} \de r\,,
\end{equation}
where $t_{\rm coll}(r,z)$ is determined by the cold-accretion recipe of Eqs. 8, 9 and 10.
We denote this cold gas spheroidal component ('gaseous bulge'), which act as a reservoir for star formation, by $M_{b,gas}(t)$,  and stress that besides dissipative collapse, it can also grow through merger events and disc instabilities.
Also, our model includes other two components:
a spheroidal stellar component ('stellar bulge') $M_{b,star}(t)$ and a low angular momentum cold gas reservoir $M_{res}(t)$, which acts as a source of material eligible to accrete onto a central black hole.

For simplicity\footnote{We need to make an assumption about the reservoir geometry because that is needed to calculate the velocity of the composite system $V_c$ [needed \textit{e.g.} in Eqs.~27 and~32, the adiabatic halo contraction factor $\Gamma$ [Eqs.~28 and~29] and the gravitational potential $\phi$
of the composite system appearing in Eqs.~15 and~34. However, the geometry of the reservoir is not expected to have a major impact on our results, given its small size relative to the other components.}, we assume that the reservoir can be described by an exponential disc surface density
\begin{equation}
    \Sigma_{\rm res}(r,z) = \Sigma_{0}(z) {\rm e}^{-r/r_{\rm res}(z)}\,,
\end{equation}
with the scale radius $r_{\rm res}$ being proportional to the influence radius of the SMBH ($r_{\rm res}=\alpha G M_{_{\rm SMBH}}/V_{\rm vir}^2$, with $\alpha\approx100$). Also, since we have no \emph{a priori} information about the geometric distribution of baryonic matter within the bulge system, and since the dynamics and thus evolution of disc structure is correlated to the mass and geometry of the bulge structure, we assume that the bulge stellar and gaseous masses settle into a Hernquist density profile
\begin{equation}
    \rho_{b}^{*}(r)= \frac{M^{*}_{b}}{2 \pi}\frac{r_{b}}{r(r+r_{b})^{3}}\,,\quad *=\mbox{stars, gas}\,,
\end{equation}
where the scale radius of this profile is related to the half light radius by $r_{b}=1.8152 R_{\rm eff}$. Using the fitting of Shen et al. (2003), we take the parametrization as a function of bulge mass to be
\[  {\log(R_{\rm eff})= }   \left\{
\begin{array}{ll}
    -5.54 + 0.56\log(M_{b})       & \quad \mbox{[ $\log(M_{b}) > 10.3$]} \\
    -1.21 + 0.14 \log(M_{b})    & \quad \mbox{[ $\log(M_{b}) \leq 10.3$]}
\end{array}\right. \]

The star formation rate (SFR) per annulus in the gaseous bulge may be computed as
\begin{equation}
   \frac{\de{\psi}_{b}}{\de r}(r,t)= 4\pi r^2 \frac{\rho_{b,gas}(r)}{t_{gas}(r)}\,,
\end{equation}
where $t_{gas}(r)$ is the dynamical time for the gas in the bulge. Therefore in order to compute the total SFR we must integrate this expression over all radii.

Energetic feedback due to supernova events may transfer significant energy into the cold ISM, causing it to be re-heated and ejected from the system. Therefore, by considering energy balance in the ISM, supernovae feedback is able to remove gas from the bulge at a rate:
\begin{equation}\label{snfb1}
    \dot{M}_{b,gas}^{SN}(t) = -\int\frac{\epsilon_{SN,b}E_{SN}\eta_{SN}\de\psi_{b}(r,t)/\de r}{\phi(r,t)}\mbox{d}r\,,
\end{equation}
where $\eta_{SN}$ is the number of Type II supernovae expected per solar mass of stars formed\footnote{A Romano IMF $\phi(m_{\star})$ is adopted: $\phi(m_{\star})\propto m_{\star}^{-1.25}$ for $m_{\star}\geq
M_{\odot}$ and $\phi(m_{\star})\propto m_{\star}^{-0.4}$ for
$m_{\star}\leq M_{\odot}$. This gives $\eta_{SN}=5\times10^{-3} M_\odot^{-1}$.}, $E_{SN}$ is the kinetic energy released per supernova event, and $\epsilon_{SN,b}$ is the efficiency of supernovae energy transfer used to remove the cold gas. Finally, $\phi(r,t)$ is the gravitational potential of the composite system (bulge, reservoir, disc and DM). Therefore, using this prescription, we see that supernovae feedback is particularly efficient in smaller halos with shallower gravitational potential wells but relatively inefficient in larger halos. Using scaling relations, Granato et al. (2000) were able to show that stars form faster in larger systems, thus exhibiting the observed 'anti-hierarchical' behavior of spheroid galaxies.

A growing body of evidence is now showing that the evolution of both the SFR within spheroids and the fueling of SMBH's are proportional to one another (Haiman, Ciotti \& Ostriker, 2004). A proposed mechanism to account for this phenomenon has been discussed (Kawakatu, Umemura \& Mori, 2003): Radiation drag due to stellar radiation may result in the loss of angular momentum at a rate well approximated within a clumpy ISM by:
\begin{equation}
    \frac{\mbox{d}\ln(J)}{\mbox{d}t} \approx \frac{L_{\rm sph}}{c^{2}M_{b,gas}}(1-e^{-\tau})\,,
\end{equation}
where $L_{\rm sph}$ is the total stellar luminosity and $\tau$ is the effective optical depth of the spheroid ($\tau=\bar{\tau}N_{int}$ where $\bar{\tau}$ is the average optical depth and $N_{int}$ is the average number of clouds intersecting a light path). Upon loss of angular momentum this gas may flow into the nuclear region, generating the reservoir of low-J material which fuels BH growth at the rate  (Granato et al., 2004)
\begin{equation}
    \dot{M}_{res} \approx  1.2 \times 10^{-3} \psi_{b}(t)(1-e^{-\tau}) M_{\odot}\,\mbox{yr}^{-1}\,.
\end{equation}
We note that, within this work we assume that $\tau$ is constant for simplicity, which allows one to rewrite Eq. 17 as
\begin{equation}
\dot{M}_{res}= A_{res}\psi_{b}(t)\,,
\end{equation}
where $A_{res}$ is a free parameter. The cold gas stored in the reservoir is then expected to accrete onto the central SMBH with an accretion rate
\begin{equation}
    \dot{M}_{bh} = \min[ \dot{M}_{visc}, \dot{M}_{edd}  ].
\end{equation}
In this formula, the viscous accretion rate is given by (Granato et al., 2004)
\begin{equation}
    \dot{M}_{visc} = k_{acc} \frac{\sigma^{3}}{G} \left( \frac{M_{res}}{M_{bh}} \right)\,,
\end{equation}
where $k_{acc}\approx10^{-2}$, whilst the Eddington accretion rate is simply $\dot{M}_{edd}=L_{edd}/\eta c^{2}$, with\footnote{This value of $\eta$ corresponds to rapidly spinning SMBH with spin parameter $a\approx0.9$ (Bardeen, 1970).} $\eta\approx0.15$ and
\begin{equation}
    L_{Edd} \approx 1.26 \times 10^{46}\frac{M_{BH}(t)}{10^{8}M_{\odot}}\mbox{erg s}^{-1}.
\end{equation}

QSO activity affects the interstellar medium of the host galaxy and also the surrounding intergalactic medium through both radiative heating and the kinetic energy input through gas outflows. Assuming that a fraction $f_{h}$ (which we treat as a free parameter) of the SMBH luminosity $L_h$ is transferred into the cold and hot gas phases, it is possible to compute the amount of cold and hot gas which is removed from the hot gas and gasoeus bulge phases as in Granato et al., 2004:
\begin{eqnarray}
            \dot{M}_{b,gas}^{QSO}&=&f_{h}\frac{2}{3} \frac{L_{h}}{\sigma^{2}} \frac{M_{b,gas}}{M_{hot} + M_{b,gas}}\,,\\
            \dot{M}_{hot}^{QSO}&=&f_{h}\frac{2}{3} \frac{L_{h}}{\sigma^{2}} \frac{M_{hot}}{M_{hot} + M_{b,gas}}\,,
\end{eqnarray}
where $\sigma=\,0.65 V_{\rm vir}$. This material is assumed to be ejected from the system.

For the chemical evolution of the cold bulge gas, we use the simple instant-recycling approximation (IRA),
whereby a fraction of mass is instantly returned into the cold gas phase in the form of processed material\footnote{We adopt a Romano et al., 2005 IMF, which has the standard Salpeter slope $1.25$ in the high mass
tail, and flattens to a slope $0.4$ below $1\, M_{\odot}$. As
shown in Romano et al. (2005), this performs better than the
Salpeter one in reproducing the detailed chemical properties of
elliptical galaxies.}.
In particular, this implies that the
effective SFR which enters the evolution equations for the gas and star bulge masses is given by
\begin{equation}
\dot{M}_b^{\rm SFR}(t)= (1-R) \int\frac{\psi_b}{\de r} (r,t) \de r\,,
\end{equation}
with $R=0.25$.
Also, we assume that $M_{hydrogen} = 0.71 M_{cold}$, where the factor takes into account the contribution of Helium and
other heavier elements.

\begin{figure}
\includegraphics[height=85mm, angle=90]{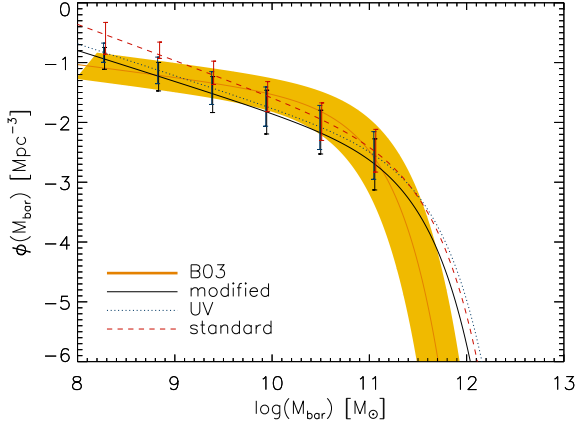}
\caption{Baryonic mass function output by our three model realisations as compared to the Bell et al. 2003b estimates. As is shown, the 'standard' approach over predicts significantly the amount of baryonic material at all masses, whereas adding the effects of a UV background and then a two-phase ISM results in progressive improvements. The most sophisticated realisation results in a good fit to the observational mass function across all observable masses. Note that the shaded area represents the $1-\sigma$ errors, discussed in fig.4. Model error-bars represent poissonian uncertanties due to the finite sample size of synthetic galaxies.}
\label{GBMF}
\end{figure}

\subsection{ Dissipationless gas collapse }

By assuming that material may collapse to form a cool gaseous disc structure during the slow accretion phase, we may add material to the disc structure at a rate which is given, as in the spheroidal case, by
\begin{equation}
    \dot{M}_{\rm coll}(z) = 4 \pi \int_{0}^{r_{\rm vir}(z)} \frac{r^{2} \rho_{\rm hot}(r,z)}{t_{\rm coll}(r,z)} \de r\,,
\end{equation}
where $t_{\rm coll}(r,z)$ is determined again by the cold-accretion determinations (see Eqs. 8, 9 and 10). If we assume a dissipationless collapse of material upon cooling within the dark matter halo, we may relate the dark halo virial radius and spin parameter to the forming disc scale radius. In particular, if we assume an exponential disc surface density profile for the stellar and gaseous components,
\begin{equation}
    \Sigma^{*}_{\rm d}(r,z) = \Sigma^{*}_{0}(z) {\rm e}^{-r/r_{d}(z)}\,,\quad *=\mbox{stars, gas}\,,
\end{equation}
the disc scale radius $r_{d}(z)$ evolves according to the scaling $r_{d}(z) = (2\pi)^{-1/2} (j_{d}/m_{d}) \lambda  r_{\rm vir}(z)f(c)^{-1/2} f_{r}(\lambda, c, m_{d} j_{d})$. The function $f(\lambda, c, m_{d} j_{d})$ may be exactly determined through (Mo, Mao \& White, 1998)
\begin{equation}\label{f_func}
    f(\lambda, c, m_{d}, j_{d}) = 2 \left[ \int_{0}^{ \infty } e^{-u}u^{2}\frac{V_{c}(r_{d}u)}{V_{c}(r_{\rm vir})}\de u \right]^{-1}\,,
\end{equation}
where $V_c(r)$ is the velocity profile of the composite system (bulge, reservoir, disc and DM) and where $m_d$ and $j_d$ are the ratios between the total mass and angular momentum of the disc component and the DM halo mass. More specifically, we take $m_d = (M_{\rm d}^{\rm stars}+M_{\rm d}^{\rm gas})/M_{\rm vir}$, and we assume $j_d=m_d$ (Mo, Mao \& White, 1998).
In order to account for adiabatic halo response, we take the standard prescription of Blumenthal (1986). In particular, denoting by $M_X(r)$ the mass of the a given component '$X$' enclosed by a radius $r$, from the angular momentum conservation one obtains
\begin{equation}\label{contraction1}
   M_{i}(r_{i})r_{i} = M_{f}(r_{f})r_{f}\,,
\end{equation}
where $r_i$ and $r_f$ are respectively the initial and final radius of the shell under consideration, the initial mass distribution $M_{i}(r_{i})$ is simply given by the NFW density profile, while $M_{f}(r_{f})$ is the final mass distribution. Also, mass conservation easily gives
\begin{eqnarray}\label{contraction2}
  && M_{f}(r_{f}) = M_{d}(r_{f})+M_{b}(r_f)+M_{DM}(r_{f}) +M_{\rm res}(r_{f})=\nonumber \\
  &&  M_{d}(r_{f})+M_{b}(r_f) +M_{\rm res}(r_{f}) + (1-f_{gal})M_{i}(r_{i})\,,
\end{eqnarray}
where $f_{gal}=M_{gal}/M_{\rm vir}$ (with $M_{gal}=M_d+M_b+M_{\rm res}$). By assuming spherical collapse without shell crossing, one can adopt the ansatz $r_{f}=\Gamma r_{i}$, with $\Gamma=\mbox{const}$ (Blumenthal, 1986), and Eqs.~28 and~29 may be solved numerically for the contraction factor $\Gamma$.

When the surface density of the gaseous disc increases, the cold gas becomes available to form stars.
However, at present, star formation is poorly understood from both a microscopic, and large-scales.
Therefore we parameterise the star formation using an empirical \emph{Schmidt law} (Kennicutt, 1998) whereby the star
formation rate is related to the surface density of cold disc gas:
\begin{equation}
    \dot{\Sigma}_{\rm sfr}(r,z) = \epsilon_{\rm sf} \left[\frac{\Sigma^{\rm gas}_{\rm d}(r,z)}{M_\odot \mbox{pc}^{-2}}\right]^{n}
M_\odot \mbox{kpc}^{-2} \mbox{yr}^{-1}\,,
\end{equation}
where $\epsilon_{\rm sf}=2.5\times 10^{-4}$ controls the star formation efficiency and $n=1.4$ is fixed to match the properties of spiral galaxies. This law is empirically proven to hold
over many orders of magnitude of gas surface density, but was shown to break down at large radii (Kennicutt, 1998)
where the surface density drops below a critical value, roughly corresponding to the Toomre (1964) stability criterion. We take this critical surface density to be
\begin{equation}
    \Sigma_{c}(r)=\frac{\sigma_{g}\kappa(r)}{3.36 G Q},
\end{equation}
Toomre, 1964. Where $ \sigma_{g}=6\mbox{km s}^{-1}$ is the velocity dispersion of the gas, $Q=1.5$ is a dimensionless constant and $\kappa(r)$ is the epicyclic frequency, given by
\begin{equation}
    \kappa(r)=\sqrt{2} \frac{V_{c}(r)}{r} \left( 1+ \frac{r}{V_{c}(r)} \frac{dV_{c}(r)}{dr}  \right)^{1/2}.
\end{equation}
Therefore, the conversion rate of gas mass to stellar mass is computed as
\begin{equation}
   \psi_d(z) = 2 \pi \epsilon_{\rm sf} \int _{0} ^{r_c} r {\Sigma}_{\rm d,\,gas}^{n}(r,z) \de r,
\end{equation}
where $r_{c}$ may be calculated by solving $\Sigma_{\rm d}^{\rm gas}(r_{c},z) = \Sigma_{c}(r_c)$ for $r_c$.
%

In order to account for feedback due to supernovae events, we may compute the amount of cold gas which is ejected from the system at each disc radius. In order to remove this cold gas from the disc, the supernovae feedback must be sufficient to unbind it, therefore we compare the amount of energy released through supernovae events at each disc radius with the
binding energy at the same radius:
\begin{equation}\label{snfb2}
        \dot{\Sigma}_{\rm SN}(r,z) = -\frac{\epsilon_{\rm SN,d} E_{\rm SN} \eta_{SN}\dot{\Sigma}_{\rm sfr}(r,z)}{\phi(r,z)},
\end{equation}
where $\phi(r,z)$ again, is the binding energy of the composite system (bulge, disc, reservoir and DM). The total amount of cold gas ejected from the system is then given by
\begin{equation}
     \dot{M}^d_{\rm SN}(z) = 2 \pi \int_{0}^{r_{\rm vir}} r \dot{\Sigma}_{\rm SN}(r,z) \de r
\end{equation}

For the chemical evolution of the cold disc gas, we use
again the IRA approximation, and the
'effective' SFR which enter the evolution equations for the gas and star disc masses is given by
\begin{equation}
\dot{M}^d_{\rm SFR}(z)= (1-R) \psi_d(z)\,,
\end{equation}
with $R=0.25$.
\begin{figure}
\includegraphics[height=85mm, angle=90]{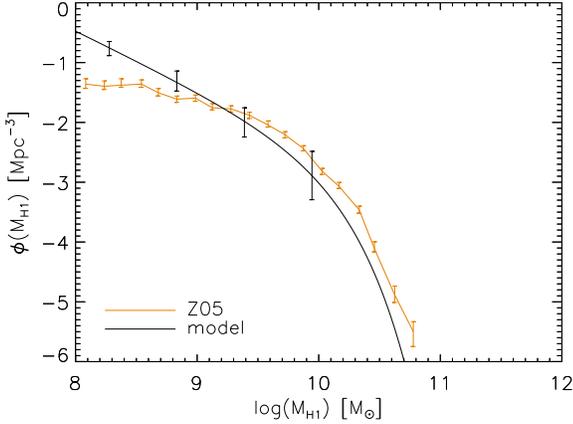}
\caption{The \hi mass function produced by our model as compared to the Zwaan et al. 2005 HIPASS galaxy sample. Using the combined effects of a UV background photo-ionizing radiation and a two-phase ISM we suppress the formation of low mass systems. We find that we reproduce both the normalisation and high mass cutoff accurately, but still slightly overproduce the number systems with $M_{\rm HI}<10^{9}M_{\odot}$, a minor effect discussed in \S4 \& 5. Observational error bars are associated to $1-\sigma$ uncertanties in determination of \hi masses. Model error-bars represent poissonian uncertanties due to the finite sample size of synthetic galaxies.}
\label{H1_mf}
\end{figure}
Also, we assume again that $M_{hydrogen} = 0.71 M_{cold}$.

Finally, it is known that when discs become self-gravitating they are likely to develop bar instabilities, get disrupted and transfer material to the spheroidal component (Christodoulou, Shlosman \& Tohline, 1995). We therefore assume that a stellar or gaseous disk is stable if
\begin{equation}\label{stability}
\frac{V_{\rm c}(2.2 r_d)}{(GM^*_{\rm disc}/r_d)^{1/2}} > \alpha_{\rm crit}^{*}\,\quad *={\rm stars\,, gas}\,,
\end{equation}
where $\alpha^{\rm stars}_{\rm crit}=1.1$ and $\alpha^{\rm gas}_{\rm crit}=0.9$ [see (Mo, Mao \& White, 1998) and references therein]. If we find that discs become unstable, we assume they get disrupted in a dynamical time and transfer their material (either stars or gas) to the bulge components.

\subsection{Improved star formation law}

Assuming that star formation may only take place inside dense molecular clouds several authors have shown that the Schmidt-Kennicutt star formation law (Eq. 30) may be reproduced within large mass systems by assuming that the star formation rate is proportional to the molecular cloud mass (Blitz \& Rosolowski, 2006, Dutton \& van den Bosch, 2009), thus:

\begin{equation}
   \dot{\Sigma}_{\rm sfr} = \tilde{\epsilon}_{sf} \Sigma_{\rm mol, HCN}
\end{equation}

Where $\tilde{\epsilon}_{sf}=13 \mbox{Gyr}^{-1}$ and $\Sigma_{\rm mol,HCN}=f_{\rm mol}R_{\rm HCN}$ is the molecular mass surface density as traced by HCN (see Gao \& Solomon, 2004, Wu et al. 2005). Thus, calculating the ratio of molecular gas to atomic gas allows us to compute the star formation rate at all scales. The fraction of gas in discs which is molecular has been extensively analysed, and shown to be closely related to the mid-plane pressure within discs (Blitz \& Rosolowsky, 2006), given by:

\begin{equation}
 P_{\rm mp} = \frac{\pi}{2}G \Sigma_{g} \left(   \Sigma_{g}  + ( \sigma_{g}/\sigma_{s} ) \Sigma_{s} \right)
\end{equation}

Where, following the detailed prescriptions of Dutton \& van den Bosch, 2009, assuming a constant $ \sigma_{g}/\sigma_{s} = 0.1$. Relating the mid-plane pressure to the formation of molecular clouds yields:

\begin{equation}
R_{\rm mol} = \frac{\Sigma_{\rm mol}}{\Sigma_{\rm atom}}= \left(   \frac{P_{\rm mp}/k}{4.3\times10^{4}}     \right)^{0.92}
\end{equation}

Thus, the molecular fraction is given by $f_{\rm mol} = R_{\rm mol}/(R_{\rm mol}+1)$, in order to relate this to the HCN fraction, we must further compute the fitting relation of Blitz \& Rosolowski, 2006:

\begin{equation}
R_{\rm HCN} = 0.1*(1+\Sigma_{\rm mol}/(200 M_{\odot}{\rm pc}^{-2}))^{0.4}
\end{equation}

Therefore, we find that, in the high mass (and thus density) galaxies, where the molecular fraction is typically $\approx 1$, we recover the standard Schmidt-Kennicutt star formation power law, with exponent $1.4$, whereas in the low-density galaxies we asymptote towards an exponent of $2.84$, suppressing star formation in these systems, in accordance with observations (see Dutton \& van den Bosch, 2009 and Blitz \& Rosolowski, 2006 for a detailed description).
Thus, throughout the evolution of each galaxy, we partition the ISM into \hi and \hii components using the above relations, and compute the star formation law (and therefore supernovae feedback) using equation 42 in order to self-consistently model each galaxy under this improved star formation law.

\begin{figure}
\includegraphics[height=85mm, angle=90]{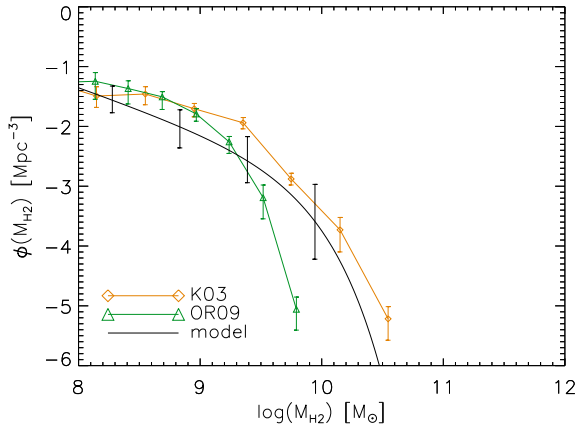}
\caption{The \hii mass function produced by our model as compared to the estimations by Keres, 2003. We find that we reproduce both the normalisation, low mass, and high mass cutoff within the observational range. Observational error bars are associated to $1-\sigma$ uncertanties in determination of \hi masses. Model error-bars represent poissonian uncertanties due to the finite sample size of synthetic galaxies.}
\label{H1_mf}
\end{figure}

\section[]{The \hi mass relationships}

In order to investigate the relationship between the stellar and the gas component in late-type galaxies,
we follow the procedure already exploited by Shankar et al. (2006), and discussed in detail in Evoli et al., 2009. We defer the reader to these papers for a more detailed discussion and highlight the main results here.
If two galaxy properties, $q$ and $p$, obey a one-to-one relationship, we can write

\begin{equation}
\phi(p) \frac{\de p}{\de q} \de q = \psi (q) \de q
\end{equation}

where $\psi(q)$ is the number density of galaxies with measured property between $q$ and $q + \de q$
and $\phi(p)$ is the corresponding number density for the variable $p$. The solution is based on a
numerical scheme that imposes that the number of galaxies with $q$ above a certain value $\bar{q}$
must be equal to the number of galaxies with $p$ above $\bar{p}$, i.e.,

\begin{equation}
\int_{\bar{p}}^{\infty} \phi(p) \de p = \int_{\bar{q}}^{\infty} \psi(q) \de q
\end{equation}

In the following the \hi mass function (HIMF) is given by; $p = \mhi$ and $\phi(p) = \mathrm{HIMF}(\mhi)$, while the variable $q$ is the
stellar mass ($M_*$) and $\psi(q)$ the corresponding Galactic Stellar mass function
(${\rm GSMF}(M_*)$). 

Using the \hi Parkes All-Sky Survey HIPASS (Meyer et al. 2004) it has become possible to map the
distribution of \hi in the nearby Universe. 
The HIMF has been fitted with a Schecter function with a power index of $\alpha = -1.37 \pm 0.03$,
a characteristic mass of $\log ( M_{HI}^*/M_\odot ) = 9.80 \pm 0.03 h_{75}^{-2}$ and a normalization
of $\theta^* = (6.0 \pm 0.8) \times 10^{-3} h_{75}^3$~Mpc$^{-3}$~dex$^{-1}$. 
Nakamura et al. (2003) estimated the LF in the $r^*$ band for early- and late- type galaxies separately.
Shankar et al. (2006) used the results of Zucca et al. (1997) and Loveday (1998) to extend these results to lower
luminosities and giving a good fit for the late-type galaxy LF in the range $10^7 L_\odot < L_r < 3 \cdot 10^{11} L_\odot$.
The GSMF, holding over the mass range $10^8 < M_* < 10^{12}$ is compared with the one of Bell (2003a),
and is in good agreement within the uncertainties due to the MLR, estimated to be around 30\%.

The correlation of galaxy properties with the halo mass ($M_h$) is extremely relevant in the framework of galaxy formation theories.
To constrain such a relation we used the association between the stellar mass and the host halo mass derived
in Shankar et al., 2006 (Eqn.12 \& Fig.1 therein), which has been obtained with the method already mentioned
but using either for the stellar mass function and the galactic halo mass function a fitting of the
observations/simulation for all the galaxies. Finding relations between each of the mas components within galaxies is of great importance, because it allows for constraints on the direct outputs from models, not requiring significant post processing, and providing useful links between the dark and luminous components of galaxies. We find that the results of the numerical outputs may be well approximated by the following analytic fitting functions:

\begin{equation}
\frac{M_{HI}}{2.38 \times 10^{8} \msun} = \frac {(\mstar / 6.1 \times 10^{7} \msun)^{2.37}} {1+(\mstar / 6.1 \times 10^{7} \msun)^{1.81}}
\end{equation}

\begin{equation}
\frac{M_{HI}}{6.07 \times 10^{8} \msun} = \frac {(M_{\rm{vir}} / 5.7 \times 10^{10} \msun)^{5.82}} {1+(M_{\rm{vir}} / 5.7 \times 10^{10} \msun)^{4.76}}
\end{equation}

Fig.2 shows the ratio of the baryonic mass components (stellar, gas and total) to the initial baryon mass associated with each DM halo obtained using the above-mentioned methods (and explained in Evoli et al., 2009), illustrating the inefficiency of galaxies, especially those of low halo mass, in retaining baryons. These relationships can be used to tightly constrain theoretical galaxy formation models in order to interpret the physical processes relevant to shape these relations.

\section[]{RESULTS}

Throughout this section we present the results of model realisations, each comprising of a sample of $\approx 10^3$ galaxies in logarithmic virial mass increments in the mass range $9.5<log(M_{v}(z=0)/M_{\odot})<13.5$ in order to encompass the observational range of galaxy-hosting DM halos. In order to make statistical predictions with each generated galaxy sample at $z=0$ we weight each DM halo with the GHMF, which mirrors the Sheth \& Tormen (2002) mass function within most galaxy sized DM haloes, but is derived in order to account for the increasing probability of multiple galaxy occupation in the highest mass haloes (see Shankar et al. 2006). Throughout this section we also plot the results for 'early' and 'late' type galaxies, by parting the populations into $[M_{bulge}/M_{total}]>0.5$ and $[M_{bulge}/M_{total}]<0.5$ respectively.

\subsection{Component evolution}

In order to illustrate the general behavior of our galaxy formation framework, Fig.3 highlights the evolution of each mass component from $z=8$ up to the present epoch. In the left-hand, center and right-hand panels we show a low-mass, intermediate mass and high-mass galaxy respectively, in order to highlight typical model outputs for each system. As can be seen, the evolution of each galaxy differs significantly due both to scatter at each mass through monte carlo selected mass accretion histories, and to the relative differences in efficiencies of the competing processes of infall, star formation and feedback on different mass-scales.

We find, in broad agreement with observations, that a typical low-mass galaxy (with $M_{v}(z=0) \approx 10^{10}$ \msolar) supports the growth of a disc structure from high redshifts showing extended star formation up to $z=0$ resulting in a gas-rich disc dominated galaxy with a negligible bulge component. In intermediate mass, \lsun galaxies, at high-redshifts gaseous collapse onto a spheroid system results in the co-evolution of SMBH's and the spheroidal component resulting in a gas poor, \emph{'red and dead'} spheroidal stellar component which acts as the bulge within the resultant formed galaxy as the disc component grows steadily from $z \leq 1.5$. The high mass galaxy shows rapid early growth of the SMBH and spheroid component, followed by an epoch of dormancy in the star formation, no significant disc structure may form in these systems and ongoing star formation is prevented due to the presence of a large SMBH which acts to effectively expel any residual gas which may infall later, only allowing for extremely mild star formation following the main growth phase. The resulting galaxy spheroidal component thus comprises an old stellar population, with little gas and negligible star formation, and the disc component is over an order of magnitude smaller than the bulge, and gas rich, showing little star formation.

These results are consistent with the evolutionary histories tuned to match the chemical properties of local galaxies of different morphological type and different host halo masses (Calura, Pipino \& Matteucci, 2008 \& references therein) which have also been shown to be consistent with photometry within the local Universe (Schurer et al. 2009) and are generally in agreement with observations of statistical samples of galaxies (Driver, et al. 2006), however, as can also be seen, due to the stochastic nature of the model, fluctuations in galaxy properties are expected, and thus we must revert to statistically representative samples of galaxy populations in order to make more robust comparisons.

\begin{figure}
\includegraphics[height=85mm, angle=90]{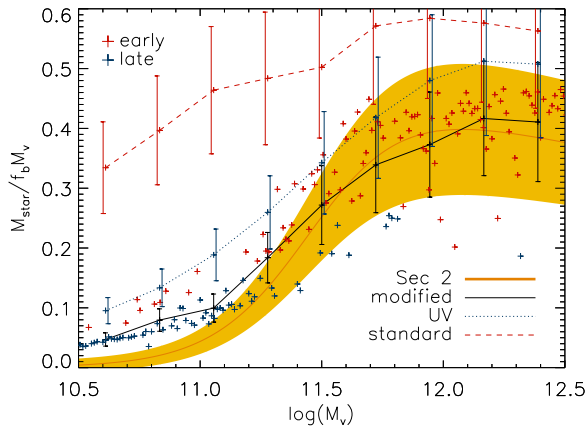}
\caption{Deriving the relationships in \S3 we show the baryon budget in the form of stellar material in DM halos of different mass. Only by exploiting the effects of a UV background and a two-phase ISM can we reproduce the global form of this relationship, finding good agreements to observations at all masses, with a slight over-prediction only in the lowest mass halos. As can be seen, 'standard' approaches significantly overproduce stellar material on all mass ranges, but the discrepancy is clearest in the low mass systems. The shaded region represents the $1-\sigma$ observational uncertanty. Model error-bars represent poissonian uncertanties due to the finite sample size of synthetic galaxies.}
\label{comparison_star}
\end{figure}

Within these systems in-falling baryonic material is shock heated and therefore may form a static atmosphere of hot-gas. Feedback from a growing SMBH may thus halt the cooling of this gas, quenching star formation and sweeping out
the ISM (See Granato et al., 2004).
Within these systems it is thought that baryonic material is shock heated and therefore cooling and feedback from a growing SMBH may quench star formation by sweeping out the ISM (see Granato et al. 2004)

\subsection{Stellar and Baryon mass functions}


The local stellar and baryonic mass functions provide a powerful constraint on theoretical models of galaxy formation: Encompassing much of the relevant physical processes which determine the assembly of baryons within DM halos. 

In order for models to reproduce observational results, it has become clear that physical processes of gas accretion, supernovae feedback and photo-ionization are most effective in the lowest mass DM halos where the shallow potential wells are inefficient at trapping and holding baryonic material. Thus these processes drive the evolution of the faint-end slope of the mass functions (Benson et al., 2002), whereas the brightest galaxies (above \lstar) are embedded within large DM halos which effectively trap baryonic material. Within these systems in-falling baryonic material is shock heated and therefore may form a static atmosphere of hot-gas. Feedback from a growing SMBH may thus halt the cooling of this gas, quenching star formation and sweeping out the ISM (See Granato et al., 2004). Coupled with the increasing subhalo contribution (through the cluster mass function) whereby in DM halos with $M_{v}(z=0) > 10^{13}$ the probability of a single galaxy occupation is low, gives rise to the relatively sharp cutoff of stellar mass within the largest DM halos (see Shankar et al. 2006 \& Somerville et al. 2008, \& contained references).

Rather than computing the spectral energy distribution (SED) assuming any mass-to-light ratio for each model galaxy (relying upon further model assumptions such as the dust-to-gas ratio, molecular cloud structure and optical depth etc) by comparing the stellar and total baryon budgets within each DM halo, we provide the most direct analysis of model outputs. However, we also note that uncertainties in the observational conversion of luminosity to stellar (and total baryonic) mass are systematically related to the spectral energy SED fitting methods used in order to extract physical quantities from the multi-wavelength observations, and on the quality of the observations themselves, this is known to have large uncertanties and we hope in a subsequent work, to utilize synthetic spectra using the detailed star formation histories, galaxy geometries, gas and dust content, in order to self-consistently model the multi-wavelength SED and make comparisons with the luminous properties of galaxies. In Fig.4 we show the Schecter function fit to the Bell et al. (2003a) estimate for the local stellar mass function, in order to generate this mass function the authors utilize a large sample of galaxies from the \emph{Two Micron All Sky Survey} (2MASS) and the \emph{Sloan Digital Sky Survey} (SDSS) converting galaxy luminosity into stellar mass using simple models to convert the optical and near infrared observations into stellar masses.


As we have shown in Fig.4, using three levels of sophistication we are able to highlight the differences between; a 'standard' model, whereby we ignore the effects of a photo-ionizing background and employ a standard Schmidt-Kennicutt (1998) star formation law, a 'UV' model whereby we include the suppression of mass flowing into galaxies through photo-ionization (see Eqn. 6 \& 7), and a 'sophisticated' model, which combines the effects of a photo-ionizing background with a two-phase ISM and modified star formation law. We will refer to these model names throughout the next sections.

More specifically, using a 'standard' model we significantly over-produce the number density of low mass galaxies. This occurs even when we increase supernovae feedback efficiencies to extremely high levels (see Mo et al. 2005 for a discussion), further pushing the parameter to higher values than $\epsilon_{sn}=0.7$ would result in largely un-physical models, and it is clear that reproducing observations using this basic framework does not occur naturally\footnote{We note also that we use a universal stellar initial mass function (IMF) which has been constrained in order to reproduce the chemical properties of galaxies accurately, increased supernovae rates are achievable using a more top-heavy IMF, however this has a strong chance to offset other properties of the formed galaxies with respect to observations (see Romano et al. 2005 for a discussion.)}. Secondly, increasing the supernovae feedback efficiencies whilst lowering the star formation efficiencies simply has the effect of reducing the stellar mass but not the gaseous mass, resulting in gas fractions which \emph{increase} with increasing supernovae efficiency and thus are in disagreement with observations (see Mo et al., 2005 for a detailed discussion).

By suppressing the initial in-fall of material due to an ionizing UV background radiation, we are able to improve agreements with the mass function. However, we achieve the best reproduction of the stellar mass function through the additional reduction in star formation efficiency when employing the 'sophisticated' star formation law, which is detemined by the amount of \hii gas present in galaxy discs.

Alone, the stellar mass function provides a crucial observation for any physical galaxy formation model to reproduce, but by simultaneously comparing both the stellar and gaseous properties of galaxies, we are able to break degeneracies between gas infall and cooling, star formation and feedback processes, resulting in a significantly improved constraint on the theoretical framework. This has led previous authors to claim that the standard model, \emph{whereby low mass systems support efficient gas cooling leading to large gaseous rotationally supported discs, which then undergo mild star formation}, significantly over-predicts galaxy \hi masses when compared to observations (see Mo et al. 2005) causing serious tension between theory and observation.

In Fig.5 we compare model results with the Bell et al. (2003b) baryon mass function. We find that suppressing the infall of baryonic gas due to an ionizing UV background significantly improves agreement between observation and model output, since material is prevented from in-falling into the halos initially and thus requiring less feedback in order to gain agreement of the low stellar and gas fractions in these halos. A final, and further improvement between model and observation is achieved by partitioning the ISM into neutral and molecular gas, which has the effect of reducing the SFR efficiency preferentially in the lowest mass DM halos. We find that the best agreement between model and observation naturally results from using the most sophisticated treatment of the ISM physics, and the initial infall of gas.  We do however, still find discrepancies in the lowest mass halos ($M_{star}(z=0) < 10^{9}$), over-predicting the baryon and stellar content within these DM halos, and slightly over-producing the number density of the most massive galaxies. This issue is discussed in \S5.

However, the observations (with their large uncertainties), remain relatively insensitive to the detailed properties of the galaxies at different masses, a strong motivation for the method adopted in \S3, whereby linear scales amplify the discrepancies between models.

\begin{figure}
\includegraphics[height=85mm, angle=90]{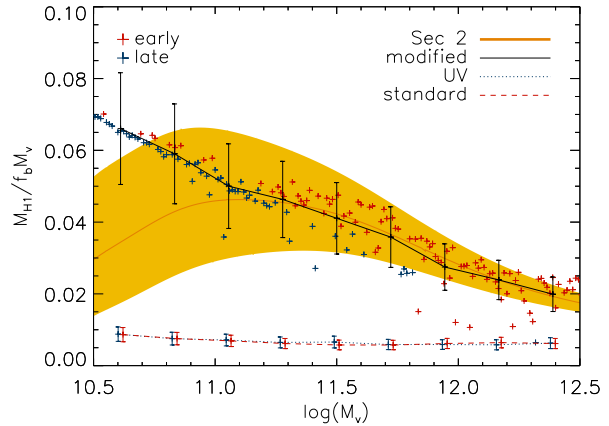}
\caption{The fraction of baryonic matter in the form of \hi relative to the cosmological fraction showing that using the 'sophisticated' model we are able to successfully reproduce the derived relationships in \S3 accurately down to relatively low halo masses, but over-predict the \hi mass by a factor of two within the lowest mass halos, we discuss the implications of this in \S5. The shaded region represents the $1-\sigma$ observational uncertanty. Model error-bars represent poissonian uncertanties due to the finite sample size of synthetic galaxies.}
\label{comparison_halo}
\end{figure}

\subsection{\hi and \hii mass functions}

Due to the implemented star formation law within the 'sophisticated' model, whereby we partition the total cold gas mass into atomic and molecular Hydrogen\footnote{Taking $M_{H_{2}}=f_{mol}M_{hydrogen}$ and  $M_{HI}=(1-f_{mol})M_{hydrogen}$}, where $f_{mol}=R_{mol/(R_{mol}+1)}$ and $R_{mol}$ is given by Eq.40 at each time-increment in order to self-consistently generate a two phase ISM, we are able to make simultaneous predictions for both the \hi and \hii mass functions. In Fig.6 we compare our model results to the observational estimates of Zwaan et al. 2005. The sample, comprising of 4315 extragalactic emission line estimated \hi masses from the HIPASS catalogue, is complete down to $M_{HI} \approx 10^{8} M_{\odot}$, which allows for an unprecedented determination of the low mass slope. We find that above $M_{HI} \approx 10^{9} M_{\odot}$ we accurately reproduce the form and normalisation of the mass function, but we do significantly over-produce the \hi masses at the low-mass extreme.

Secondly, in Fig.7 we compare the \hii mass function as derived from the CO-mass function determination by Keres et al. 2003 based on the FCRAO extragalactic CO survey of 200 galaxies (Young et al. 1995). The conversion from CO to \hii is achieved through a '$\chi$-factor' which proves to be a delicate and difficult task to determine, we utilize the \hii mass function derived by Obreschkow \& Rawlins, 2009 as this appears to be the most robust determination, and also plot the \hii mass function derived within the originalKeres et al., 2005 work (using a constant $\chi$-factor). We refer the reader to their work for further details. We find that within observational errors, we reproduce this function over the entire observational range, however, we note that the large uncertanties within the determination of the precise value of the $\chi$-factor means that we are relatively loosely constrained, and we view this result as a general prediction of our model, rather than a constrained observational match.

Physically interpreting these results, we may conclude that utilising a two-phase ISM whereby star formation scales only indirectly with the total gas mass, but directly with the molecular \hii mass, we find that we may accurately reproduce both the \hi and \hii mass functions of galaxies apart from an over-prediction of neutral \hi within the lowest mass halos. This discrepancy also appears within the stellar mass function (Fig.4) and the baryonic mass function (Fig.5), although it is greatly improved using the 'sophisticated' ISM treatment.

\subsection{Stellar-to-Dark matter properties}

Exploiting the methods outlined in \S3 we are able to make observational mappings between DM halo masses and their contained stellar mass. In Fig.8 we show the relative efficiency of conversion of baryonic material into stellar mass within DM halos relative to the cosmological baryon to dark matter ratio $f_{b} = \Omega_{b}/ \Omega_{m} = 0.16$. These relationships provide direct observational constraints onto galaxy formation models and allow for a direct analysis of the relevant processes occurring over a range of mass scales (see Shankar et al. 2006). As can be seen, the lowest mass systems $(log(M_{v}/M_{\odot})<11)$ are strongly dark matter dominated, due to the inefficiency of weak gravitational potentials in these systems being able to capture and contain baryonic material and then efficiently form stars. Typically these systems are strongly disc-dominated, late-type galaxies.

Intermediate mass systems $(11<log(M_{v}/M_{\odot})<12)$ generally contain a range of morphological types with significant spheroid and disc components. The stellar content in these systems is a strong function of the mass, whereby the larger systems become richer in stellar mass fraction reaching a 'peak' ratio of stellar to dark matter where there is a minimum efficiency in the combined effects of supernovae and nuclear feedback processes. When we approach the highest mass DM halos $(log(M_{v}/M_{\odot})>12)$, feedback through nuclear activity generated by an accreting SMBH is able to effectively remove the material. This effect becomes stronger in the higher mass systems and thus causes the flattening and slight downturn of the stellar conversion efficiency.


Over-plotting model outputs using the 'standard', 'UV' and 'sophisticated' models, we may easily differentiate between outcomes due to the increased sensitivity of the plotting ranges: We see that the models which include the effects of a UV background show significant improvements over the entire mass range, and using the two-phase ISM physics and adopting the SFR dependent on \hii mass, we are able to reproduce the observations to a high accuracy throughout the entire mass range due to the relative decrease in SF efficiency, with the only discrepancies at the lowest masses. It is important to note here that due to the sequential build-up of matter within halos from high redshifts to low, the importance of ionizing backgrounds and star formation on low mass systems manifests throughout the entire mass range at $z=0$ (see Somerville et al. 2008 for further discussion of this). Under the standard \lcdm scenario whereby small structures collapse first and form progressively larger systems, large structures at $z=0$ are therefore significantly influenced by small scale processes at higher redshifts, where their constituent parts were forming and evolving, thus the low mass behavior is of global importance to galaxy formation theory.

\begin{figure}
\includegraphics[height=85mm, angle=90]{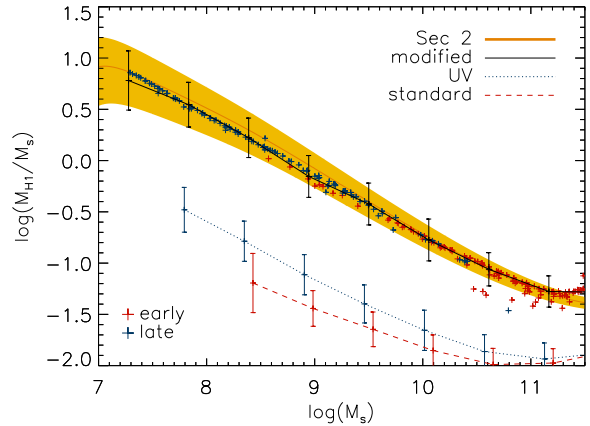}
\caption{\hi-to-stellar mass ratio as a function of the stellar mass. Using derived relationships from \S3 we are able to assess the relationship between \hi and stellar mass, finding that using a two-phase ISM and related SFR computations we  accurately reproduce the observations across the entire observational range, unlike the more simplistic frameworks, where we systematically convert too much cold gas into stellar material. The shaded region represents the $1-\sigma$ observational uncertanty. Model error-bars represent poissonian uncertanties due to the finite sample size of synthetic galaxies.}
\label{MHI_over_Mstar}
\end{figure}

\subsection{Gas-to-Dark matter properties}


In analogy to the previous section, in Fig.9 we plot the fraction of \hi to DM halo mass we are able to see that the gas fractions are highly sensitive to the different physical prescriptions, the scatter is attributed to the fact that the cold gas (and thus \hi mass) component at any time is controlled by the competing processes of infall, star formation, feedback and recycling. We also find that, due to the uncertainties and intrinsic dispersion in the relationship between the dark matter and gaseous matter, the observational trend is not as constrained for the gaseous as for the stellar component. Observationally, the lowest mass DM halos become progressively more depleted in \hi, due to suppression of infall and increased efficiency in generating gaseous outflows, intermediate mass DM halos ($11>(log(M_{v}/M_{\odot})>11.5)$) possess the most \hi mass, due to the global maxima in trapping and containing baryonic material in DM halos (or conversely, a global minima in feedback processes), and at high masses we see a progressive loss of \hi abundance in halos, signifying an increase in removal of gaseous material through cooling inefficiencies and nuclear feedback processes.

In Fig.9 we show the outputs for each of the three realisations, as can be clearly seen, the only successful reproduction of the majority of the observations is the 'sophisticated' model, whereby we model both the effects of a two-phase ISM and of an ionizing background radiation. Therefore modeling the star formation rate depending linearly on the surface density of molecular gas in the disc component results in a correct conversion of cold gas to stellar material (as can be seen from simultaneous fits to both Fig.8 \& Fig.9), using the standard Schmidt-Kennicutt (1998) star formation law (and their constrained normalisation) we convert too much material from gas into stars, resulting in an over-production of stellar mass and an underproduction of gas mass in all DM halos, whereas the modification outlined in Dutton \& van den Bosch (2009) allows for the correct fraction of virial mass in the \hi phase, apart from in the lowest mass systems ($11>(log(M_{v}/M_{\odot}$), where we find observation and model predictions in alarming disagreement.

As with previous works, but to a lesser extent, we find that we over-produce the amount of gas in the lowest mass DM halos, and more worryingly, we do not reproduce the observed down-turn in gaseous mass within the lowest mass systems $(log(M_{v}/M_{\odot})<11)$. This mismatch clearly suggests that despite our suppression of baryonic material onto low mass DM haloes, and improved star formation prescriptions we are still missing extra factors which dominate at $M_{v}(z=0)<10^{11}M_{\odot}$. We my attribute this effect with our lack of description of environmental effects, since the lowest mass halos are more probable to be located within larger over-densities and therefore subject to external forces (see the \S5 and Mo et al. 2005 for more details).

\subsection{Stellar-to-Gas properties}


It is suspected that within the lowest mass systems, gas collapse, star formation and supernovae feedback result in a self-regulated conversion of gas into stars, thus these effects can be seen to produce precisely the correct fraction of stars to gas in Fig.10. Within our model the cold gas component fluctuates more than any other galaxy component since it is constantly being replenished due to infall and stellar recycling, exhausted through star formation and expelled through feedback processes. Therefore, directly comparing each galaxies gaseous and stellar properties provides a stringent comparison between observation and model. Interestingly we find that plotting these quantities, we reproduce a tight correlation with relatively little scatter.


We find an overall agreement to the data across the entire mass range under investigation using the 'sophisticated' model, showing that the \hi fraction becomes increasingly large with decreasing stellar mass, having roughly equal stellar and \hi masses at $M_{s} \approx 10^{9}M_{\odot}$ increasing to a factor of $\approx 100$ times more \hi than stellar mass within systems with $M_{s} \approx 10^{7}M_{\odot}$. Conversely, without using the effects of a two-phase ISM and UV background radiation, we find a general offset in all masses, with an over-efficient conversion of cold gas to stars (as may also be seen in Figs.8 \& 9).

We also find that this result is relatively robust against parameter choices; only having a significant dependence on the star formation efficiency parameter (which is constrained observationally). We may attribute this to the fact that star formation is modeled as a function of the surface density of \hii in galaxies and within the low mass disc dominated systems is relatively inefficient (due to the low surface densities and relatively high star formation thresholds) resulting in a well defined conversion of cold gas to stars and hence, \hi to stellar mass ratio. This conclusion also has important consequences when considering the overproduction of both stellar and \hi masses in the lowest mass DM halos, since we accurately reproduce the self-regulation properties of galaxies (as the correct balance between star formation and feedback is required to simultaneously reproduce the correct gas and stellar mass budgets in DM haloes). Overproduction in the low mass regions within Fig.8 \& Fig.9 therefore, must be attributed to the accretion of too much material, a conclusion which is confirmed by several other authors (see Mo et al. 2005 \& references therein).

We would like to plot also the \hii mass counterparts for figures 8 \& 9, however, due to the unconstrained mass functions, as discussed in \S 4.3, we are not able to construct conclusive observational constraints. Secondly, since partitioning the ISM within the single-phase ISM models is typically something relatively ad-hoc, we do not perform this analysis here.

\begin{figure}
\includegraphics[height=85mm, angle=90]{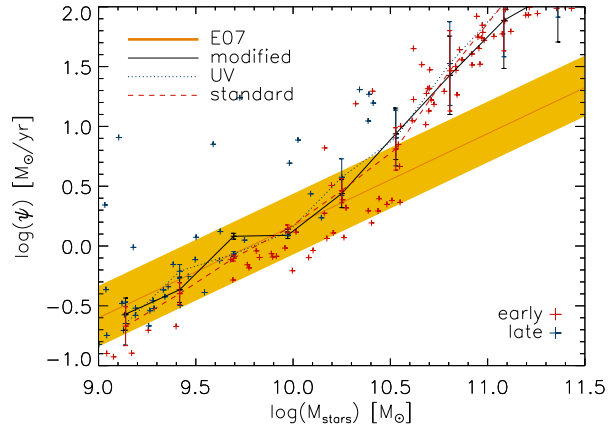}
\caption{The predicted SFR as function of the stellar mass at $z=0$ for our three realisations compared to Elbaz et al. 2007. As can me seen, the star formation rates in all three models are similar, all showing good agreements in the low/intermediate mass systems but models overpredicting the SFR in the highest mass systems. The shaded region represents the $1-\sigma$ observational uncertanty. Model error-bars represent poissonian uncertanties due to the finite sample size of synthetic galaxies.}
\label{SFR}
\end{figure}

\subsection{Star formation properties}


A final useful diagnostic to be used to constrain galaxy formation models is the instantaneous star formation rate (SFR) which is found to vary significantly within galaxies of different stellar mass. In Fig.11 we compare each model realisation with the observational SFR estimates as a function of stellar mass by Elbaz et al. 2007, who used a large sample of SDSS galaxies with spectroscopic data in order to accurately determine the SFR (see Brinchmann et al. 2004).


We find little difference between the SFR predicted by each model realisation as expected since each star formation rate prescription is constrained by $z=0$ galaxy properties. Moreover, we generally find that within the low mass, disc-dominated region ($M_{s} < 10^{10}M_{\odot}$) we obtain a good agreement between model and observation but we significantly over-predict the star formation rates in the largest galaxies. We may attribute this to a lack of AGN quenching within the disc component of the largest mass systems at late times, which has been applied in several models as a rather \emph{ad-hoc} 'radio-mode' feedback (see Croton et al. 2006, Bower et al. 2006). Since, within our models we do not have any suppression of the growth of discs around large pre-formed bulges, aside from the late transition from the spheroid formation epoch to the disc formation epoch. We hope to investigate the effects of energetic feedback from a formed SMBH-spheroid system on the late properties of the disc since we hypothesise that; despite having a negligible effect on the growth of a pre-formed SMBH, star formation is prevented within the spheroid component at late times since even arbitrarily low accretion rates onto the central SMBH results in energetic feedback capable of heating the ISM, however, within our framework, during the quiescent disc growth phase no material is assumed to collapse onto the spheroid structure, even with a slight adjustment to our model we may allow for some material to collapse at late times onto the spheroid-SMBH system, resulting in the quenching of star formation in the larger systems and naturally generating Seyfert-type galaxies. However, for simplicity we have neglected this effect within this work, and hope to investigate the physical mechanisms capable of generating this self-consistently, within a subsequent work.

\subsection{Two component ISM properties}

As a main advance of this model over current SAMs, we detail the properties of the ISM by modeling the formation of molecular clouds (\hii regions) through pressure arguments within the disc component. This enables us to modify the star formation law, and thus allows us to gain insights into the more detailed gas-properties of normal galaxies under the semi-analytical framework.

Within this work we highlight the importance of modeling the ISM in two-phases advocating it as a simple, yet important advance over current frameworks; whereby the formation of \hii regions is determined by the planar pressure within the gaseous disc structure. This added ingredient is important for two main reasons: Firstly the decline in \hii regions within low mass systems results in a higher fraction of cold gas in the form of neutral \hi which is thus detectable through conventional 21-cm line surveys (see Barnes et al. 2001), whereas \hii mass estimates prove to be significantly more difficult, relying on uncertain conversion factors between \hii and CO-lines, therefore, assuming a single conversion between total cold gas and \hi, as is commonly done in SAMs provides inaccurate outputs. Secondly, the star formation properties of galaxies have been shown to relate explicitly to the detailed properties of the internal structure of the ISM (see Krumholz et al. 2009, Gnedin et al. 2009, Obreschkow \& Rawlings, 2009), and therefore this added layer of complexity should now be embedded within current SAMs.

In Fig.12 we show the fraction of cold gas which is in the form of \hii within the disc component as a function of total stellar mass (disc and bulge masses). As can be seen, there is a tight relationship with little scatter in the lowest mass systems due to the overall dominance of disc components and the molecular fraction decreases steadily with decreasing stellar mass, we find a peak in molecular fraction corresponding to approximately $80\%$ \hii at $M_{s} \approx 10^{11}M_{\odot}$ where the disc mass reaches a maximum. Above this mass the spheroid component dominates and the disc surface density thus drops, lowering the efficiency of molecular cloud formation rapidly.

\section{Conclusions}

\begin{figure}
\includegraphics[height=85mm, angle=90]{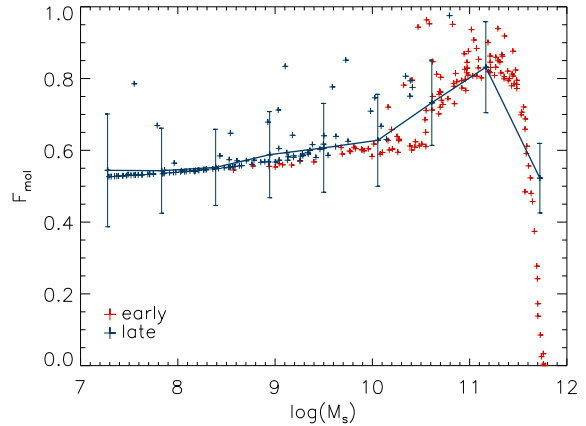}
\caption{The molecular fraction of gas within galaxy disks. Showing that in low surface density galaxies, the formation of
molecular clouds is suppressed, and thus star forming regions are diminished, and within the high-mass systems, the spheroid component dominates reducing the surface density in disks and therefore the molecular gas fraction rapidly declines. We hope to compare this result to observational studies as they become available. Model error-bars represent poissonian uncertanties due to the finite sample size of synthetic galaxies.}
\label{mol_fraction}
\end{figure}

Prompted by several theoretical attempts to model the low-mass end of the stellar mass function (Somerville et al. 2008 \& references therein) and significant observational effort to constrain the \hi mass function using large surveys of galaxies (Barnes et al. 2001, Zwaan et al. 2005), within this work we have developed a physically motivated model in order to explain the inefficiencies of low-mass DM halos in trapping baryonic material and forming stars. Motivated primarily by the importance of physical descriptions within small scale systems which form the building blocks of larger systems within a hierarchically clustered Universe, also because theoretical models have either neglected, or find significant troubles in simultaneously matching both the stellar and \hi mass functions within the low mass end (see Mo et al. 2005).

Therefore, within this paper we have utilized well constrained observations of the stellar (Bell et al. 2003a, 2003b) and \hi (Zwaan et al. 2005) mass functions and employed a numerical technique (Shankar et al. 2006) in order to derive relationships between galaxy properties and their host DM halos. Assuming a one-to-one mapping of these systems, we are thus able to make detailed comparisons between models and observations.

In order to interpret these phenomenological relationships physically, we develop a cosmologically grounded galaxy formation model outlined in C09 but with several significant modifications. Under this framework we follow the development of baryonic material as it accumulates and evolves within growing DM halos and is subject to cooling, heating, possible angular momentum losses, star formation and recycling, and feedback through both supernovae events and the growth of a central supermassive black hole. By comparing three model realisations with varying sophistication; one using a 'standard' approach, whereby the cosmological baryonic fraction $f_{b} = \Omega_{b}/\Omega_{m}$ is allowed to accrete within halos throughout their lifetimes and the star formation in discs (and thus the vast majority of low mass systems) is given by the Schmidt-Kennicutt (1998) star formation law, determined by the total cold gas mass, a 'UV' model, whereby the infall of baryonic material within low mass DM halos is suppressed due to the presence of an ionizing UV background (see Eqn.6, 7), and finally a 'sophisticated' model with a modified infall due to UV radiation and a modified star formation law which requires a two-phase ISM, comprising of neutral \hi and molecular \hii and typically results in a lower star formation rate within low mass systems at early times.

These realisations clearly demonstrate that using a 'standard' approach, the most simple case is not able to simultaneously match both the stellar and \hi mass functions, significantly over-producing low mass galaxies (see Figs.4, 5, 6, 8 \& 9), whereas the use of a 'sophisticated' approach, we are able to match observations reasonably accurately to relatively low masses, finding discrepancies within the lowest mass systems. As an additional consequence of the modified star formation law in the 'sophisticated' model realisation, we naturally partition the cold gaseous ISM into \hi and \hii components, allowing for predictions of both of these quantities without the need for further assumptions, finding that we are able to match both components simultaneously to a good accuracy over the entire mass range when comparing to the \hii results of (Keres et al. 2003, Obreschkow \& Rawlings, 2009, Fig.7), and only over-producing the amount of \hi mass in the lowest mass galaxies (see Zwaan et al. 2005 \& Fig.6).

Finally, analysing several properties of galaxies against their stellar masses, we find that, unlike the simple approaches, the \hi-to-stellar mass ratio is accurately reproduced using the sophisticated treatment (see Fig.10), indicating that the self-regulation of star formation allows for the correct conversion of material, is relatively insensitive to the infall and supernovae feedback rates within physical limits. Comparisons of the star formation rates within these models however shows little difference at $z=0$ as expected, and indicates a secondary problem with our simple physical model, over-predicting the SFR in the largest systems (with $M_{v}>10^{11}M_{\odot}$), however, these systems are typically spheroid-dominated and are therefore relatively insensitive to the details of the disc formation recipes.

It is clear to assess the limitations and the manifestations of this relatively simplistic approach to galaxy formation modeling, under our improved framework we still require large feedback efficiencies in order to sufficiently suppress star formation within the lowest mass systems: As has been studied by Mo et al. 2005, preheating through pre-virialised structure formation may further reduce the baryonic infall onto the lowest mass halos, further reducing the need for such high SN efficiencies, other environmental effects such as tidal stripping and harassment may help to further improve the theoretical framework. However, despite these further degrees of freedom, we also note that changes in the initial mass function towards something more top-heavy will naturally have a higher supernovae fraction per stellar population, this therefore further reduces the need for highly efficient supernovae feedback within these models. In order to assess these additional effects we would need to account for all the environmental effects associated with galaxy evolution, expanding our modeling from a single mass accretion history to a full merger-tree framework, however, this adds a great deal more complication and uncertain physics (such as the evolution of satellite structures within DM halos, the merger rates of galaxies, the outcome of galaxy components in mergers of different ratios), this will be the subject of a further analysis in a subsequent paper.

A second, minor shortcoming of our modeling appears to come from the complete separation of spheroid and disc growth within the two epochs of DM halo growth, whereby the disc and spheroid only share material through disc instabilities which are generally quite rare at late times due to the stabilization generated by the pre-formed spheroid. 
Despite showing that our quasi-monolithic scenario for the growth of galaxies to show promising results, we also hope to include explicitly the effects of merger events and environmental effects in a future work.

In conclusion, focusing mainly on the low mass galaxy population, adopting several theoretical improvements over 'standard' SAMs, we are able to simultaneously match both the stellar, \hi and \hii mass budgets within DM halos, and the star formation properties of galaxies within the observational ranges. This promising result indicates that at present, the 'standard' approach to modeling the low mass evolution of galaxies is somewhat over-simplified within current SAMs which only have a single-phase ISM, and, due to the hierarchical nature of structure formation, may manifest as significant tensions in progressively larger systems within a full merger-driven framework (see Somerville et al. 2008 for a discussion). We therefore advocate the use of more sophisticated treatments of the interstellar medium within current and future SAMs.

Assessing the limitations of our framework, we conclude that further suppression of infall onto the lowest mass systems would allow for a further reduction in the need for strong supernovae feedback and should further ease tensions between models and observations, this could only come through environmental effects such as tidal shocks or gravitational pre-heating (Mo et al. 2005), however this effect has not been studied in detail through hydrodynamic simulations and remains to be fully investigated. By adding a channel whereby even small amounts of gaseous material may be transferred to the spheroid component during late times, small amounts of 'radio mode' AGN activity may be triggered, little affecting the spheroid or the SMBH masses, but significantly lowering the SFR in the discs, preferentially at large masses, naturally resulting in Seyfert-type active galaxies and reducing the SFR in these large discs, hopefully bringing Fig.11 into better agreement with observations, we also hope to investigate the pan-redshift galaxy population under this framework (Cook et al., 2009b submitted).

Interestingly however, within this relatively simplistic framework we are able to self-consistently reproduce several of the key observations, it is therefore clear that mergers, to some degree, are not the dominant driver for the global evolution of the galaxy population. It will therefore provide a useful exercise to mount our physical prescriptions onto a full merger-tree DM background which should allow us to model environmental effects consistently. The main results from this paper indicate however, that using a relatively simple framework, we find an reasonable agreement to the stellar, \hi and \hii mass functions of galaxies arising naturally, thus, we advocate all current SAMs to begin to incorporate two-phase ISM physics into their frameworks.

\section*{Acknowledgments}

We thank P. Salucci for providing the initial seeds for this work, and also to F. Shankar and A. Schurer for stimulating discussions which helped the progress of this work. MC thanks L. Paulatto for considerable computational assistance, and we thank A. Ferrara for careful reading of the manuscript. MC has been supported through a Marie Curie studentship for the Sixth Framework Research and Training Network MAGPOP, contract number MRTN-CT-2004-503929. E.B. acknowledges support from NSF Grant No. PHY-0603762.

\bibliography{hi_in_galaxies}
\bibliographystyle{sources/mn2e}

\bsp

\label{lastpage}

\end{document}